\documentclass[
superscriptaddress,
amsmath,amssymb,
aps,
reprint,
physrev,
]{revtex4-2}

\usepackage{graphicx}% Include figure files
\usepackage{subcaption}
\usepackage{dcolumn}% Align table columns on decimal point
\usepackage{bm}% bold math
\usepackage{booktabs}

\begin{document}

\preprint{APS/123-QED}

\title{Cavity-QED Simulation of a Maser Beyond the Mean-Field Approximation}% Force line breaks with \\

\author{Niall Randall Carrera}
\affiliation{Department of Materials, Imperial College London, Exhibition Road, South Kensington, SW7 6AZ, UK}

\author{Yining Jiang}
\affiliation{Department of Materials, Imperial College London, Exhibition Road, South Kensington, SW7 6AZ, UK}
%\affiliation{Department of Applied Physics, Aalto University, Aalto, Espoo, Finland}

\author{Xinpeng Shu}
\affiliation{Department of Materials, Imperial College London, Exhibition Road, South Kensington, SW7 6AZ, UK}

\author{Hao Wu}
\affiliation{Center for Quantum Technology Research and Key Laboratory of Advanced Optoelectronic Quantum Architecture and Measurements (MOE),
School of Physics, Beijing Institute of Technology, Beijing 100081, China}

\author{Mark Oxborrow}
\affiliation{Department of Materials, Imperial College London, Exhibition Road, South Kensington, SW7 6AZ, UK}

\begin{abstract}

Based on the well-known Tavis-Cummings model of cavity quantum electrodynamics (QED),
we introduce a method for quantum-mechanically simulating the dynamics of experimental masers beyond the mean-field approximation (MFA)
that takes into account the spatial variation of the a.c.~magnetic field of the maser's amplified microwave mode across its gain medium. 
The distribution in the coupling between the amplified mode and the medium's very large number (typically $10^{17}$) of spatially distributed quantum emitters
can be determined straightforwardly for a given geometry and composition using an electromagnetic-field solver. Upon discretising this distribution
as a histogram over a small finite number of bins, we assign --as an approximation-- the same coupling to all emitters that
fall within the same bin, where the value of this coupling equals the center value of the bin's range.  
With our approximate Hamiltonian arranged as a weighted sum over these bins,
we generate expressions for expectation values of operators in the Heisenberg picture to second order in cumulant expansion,
using the publicly available QuantumCumulants.jl package in Julia.
For ten evenly spaced bins, our model, which can be run on a laptop computer,
is used to simulate the recorded output from an experimental maser with a pentacene-doped
\textit{para}-terphenyl gain medium.
We find that it replicates the quantum-mechanical features of the measured maser's dynamics, in particular its damped collective Rabi oscillations,
more closely than the  standard TC model under the MFA can, with an R$^2$ value of 0.77, as opposed to 0.27.
Our model should thus aid the quantitative engineering of improved, optimised maser designs.  

\end{abstract}

\maketitle

%\tableofcontents

\section{\label{sec:level1}Introduction}

Largely out of necessity, simulations of quantum-mechanical systems often incorporate a cascade of rather drastic approximations, many of which are uncontrolled. Their names include (this list is not exhaustive and in no particular order): adiabatic, Born,
rotating-wave\cite{burgarth2024}, Markovian, weak-coupling, impulse, finite-dimensional, Trotter, Holstein–Primakoff, and mean-field. This paper quantitatively assesses the validity of the last of them, as it has been applied over recent years to the simulation of optically-pumped masers. \\
\indent As its well-chosen acronym might suggest, a “ma\textbf{ser}”, the microwave equivalent of a laser, works through the \textbf{s}timulated \textbf{e}mission of electromagnetic \textbf{r}adiation from atoms, molecules, or other emitters, the constituents of
a `gain medium'\cite{townes}. To date, the deployment of masers in widespread applications has been severely hampered by their need for cryogenic and/or ultra-high vacuum conditions. Often, a controlled magnetic environment, provided by either a magnet or some form of magnetic shielding, is also required\cite{OGmaser, solidstate}. The arrival of optically-pumped solid-state masers capable of functioning at room temperature and in zero magnetic field offers a way to avoid these encumbrances. The first such maser used pentacene-doped \textit{para}-terphenyl (PTP) as an optically pumped gain medium\cite{oxborrow2012room}; this is the type of maser that we shall simulate here since the spin dynamics of PTP are accurately known\cite{wu_spindynamics,pentODMR}.\\
\begin{figure}
    \centering
    \includegraphics[width=1\linewidth]{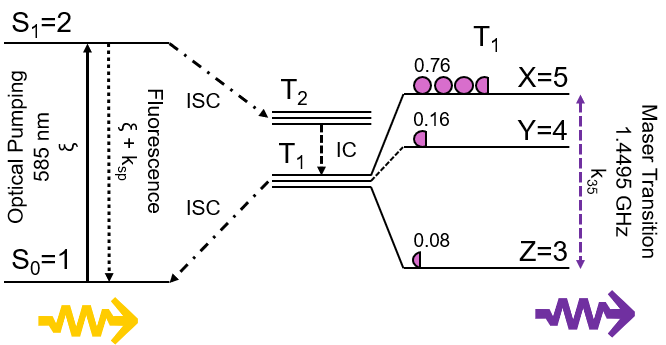}
    \caption{Pentacene molecules are driven from their singlet ground states, $S_0$ into their first excited singlet states, $S_1$, by their absorption of yellow pump light (photons) at rate $\xi$. From there, the molecules  undergo spin-selective intersystem crossing into the triplet manifold $T_2$\cite{pentacene, pentacene2}, before decaying to $T_1$ through internal conversion, which preserves the ISC-induced population inversion across the X and Z sub-levels. The frequency difference is 1.4495 GHz, corresponding to an L-band microwave photon.}
    \label{fig:jablonski}
\end{figure}
Upon optical excitation from its ground state, S$_0$ (a singlet), a molecule of pentacene will arrive at its lowest excited singlet state, S$_1$;
see Fig.~\ref{fig:jablonski}. From there, it can either fluoresce back down to S$_0$ or else undergo intersystem crossing (ISC) into pentacene's second lowest triplet state, T$_2$. The sublevels of both T$_2$ and the lowest triplet state, T$_1$ (whose energy lies lower than S$_1$'s), each comprise 3 sublevels labeled X, Y, and Z, which are split evenly at zero applied magnetic field on account of pentacene's low rotational symmetry\cite{waals01}.
ISC from S$_1$ into the X, Y, and Z sub-levels of T$_2$ occurs at different rates, resulting in an initial population distribution of 0.76:0.16:0.08, respectively, across them\cite{pentacene, pentacene2}.
These differences in population are preserved upon the molecule's rapid decay from T$_2$ down to T$_1$,
%
%Pentacene's lowest excited state, into which most molecules will fall %upon optical excitation, is a triplet, denoted as T$_1$, whose %sublevels, labeled X, Y and Z, are split even at zero applied %magnetic field on account of pentacene's low rotational %symmetry\cite{waals01}.
%Upon reaching the lowest excited singlet state, S$_1$, through %optical excitation, molecules can access these sub-levels through %intersystem crossing (ISC) from S$_1$ to the second-lowest triplet %state, T$_2$, followed by internal conversion from T$_2$ to T$_1$. %The ISC into the X, Y, and Z sub-levels of T$_2$ occurs at different %rates, resulting in an initial population distribution of %0.76:0.16:0.08, respectively\cite{pentacene, pentacene2}.
%
%Since these relative populations are preserved by internal %conversion, one can achieve a 
enabling the generation of considerable population inversion between X and Z, which differ by 1.4495 GHz in frequency\cite{pentacene, pentacene2,oxborrow2012room}. As spin-lattice relaxation in PTP is relatively slow, even at room temperature, this inversion will persist for several tens of microseconds\cite{oxborrow2012room,wu_spindynamics,wu2020room}. \\
\indent In the maser that we simulate here, the PTP gain medium resides inside a cylindrical cavity that supports a TE$_{01\delta}$ microwave mode, mechanically tuned to the same frequency as T$_1$'s X-Z transition \cite{oxborrow2012room}. Located around the gain medium is a `magnetic flux concentrator' in the form of a ring of monocrystalline strontium titanate (STO), this material being chosen for its high-permittivity and low dielectric loss. The a.c.~electric field of the TE$_{01\delta}$ mode circulates mostly inside the ring, and the a.c.~displacement current associated with the a.c.~polarised STO generates 
(through Maxwell's extension of Amp\`{e}re's law) a strong a.c.~magnetic field in the gain medium. Through magnetic-dipole interaction, the latter field induces transitions between the X and Z sub-levels of pentacene's triplet ground state, T$_1$, with a coupling strength (coefficient) equal to $g$\cite{g}. A sufficiently large $g$ allows for coherent amplification of the microwave mode\cite{lamb1964theory}. In the Tavis-Cummings model, which considers $N$ two-level atoms coupled to a common electromagnetic mode, the collective `renormalised' $g$ (for the so-called `bright sector' of the complete model's Hilbert space\cite{larson21}) scales with $\sqrt{N}$\cite{tavis1968exact, breezey}.
\indent To make a useful maser, it is often advantageous for the number of molecules, here also denoted as $N$, that interact with the microwave mode  to be as large as possible. Doing so enables the maser's cooperativity, defined as $C=4g^2/\kappa\chi$, to exceed the threshold for masing, namely $C>1$ (here $\kappa$ is the decay rate of the cavity mode and $\chi=2/T_2^*$ is the spin dephasing rate)\cite{breeze2018continuous, breezey}. A typical number of pentacene molecules resident in the gain medium is $N\approx10^{17}$\cite{wu2020room}, a fraction of which, at any one moment, will be in one of T$_1$'s three sublevels (or possibly some linear superposition thereof) or a different state.\\
\indent We point out here that the approach adopted below draws many parallels with the simulation of superradiant lasing for an inhomogeneously broadened optical gain medium undertaken by Bychek \textit{et al.} \cite{bychek2021superradiant}. In contrast to this work, here we are mainly focused on the transient photon and spin dynamics at the start up of the lasing/masing.
With regard to simulating the observed quantum-mechanical features of our maser's dynamics,
one needs to capture how an ensemble of $N$ two-level (X-Z) systems interact with a single cavity mode. 
The Tavis-Cummings Hamiltonian in the interaction picture, and with the ensemble taken to be homogeneously resonant with the cavity mode (i.e., $\omega_{mode}-\omega_{XZ}=0$ for all molecules) \cite{tavis1968exact}, provides a way of doing this:
\begin{eqnarray}
\hat{\mathcal{H}}=\sum_{j=1}^{N}\hbar g_j(\hat{a}^{\dagger}\hat{\sigma}_j^-+\hat{a}\hat{\sigma}_j^+).
\end{eqnarray}
This basic Hamiltonian is extended with jump operators to capture transitions to and from the gain medium's other energy levels.
Dephasing and cavity losses are added phenomenologically, resulting in a master equation for the (reduced) density matrix in the Schr\"{o}dinger picture:
\begin{eqnarray}
\dot{\hat{\rho}}=-\frac i\hbar[\hat{\mathcal{H}},\hat{\rho}]+\sum_{k}\gamma_k\mathcal{D}[\hat{L}_k]\hat{\rho},
\end{eqnarray}
Where $\mathcal{D}[\hat{L}_k]\hat{\rho}=\hat{L}^{\dagger}_k\hat{\mathcal{\rho}}\hat{L}_k-\frac12\{\hat{L}_k^{\dagger}\hat{L}_k,\hat{\mathcal{\rho}}\}$. Here we are immediately faced with the curse of dimensionality: to describe $N$ two-level systems alone requires a density matrix $\hat\rho$ containing $4^N$ elements, which is computationally unfeasible. Since we do not seek to know the time evolution of individual molecules (rather, only the evolution of the expectation values of certain operators or operator products involving sums over molecules), an alternative approach\cite{quantumcumulants}, formulated in the Heisenberg picture, evolves operators through time according to a quantum Langevin equation. Assuming its associated noise is white and does not contribute to averages, we arrive at\cite{quantumcumulants}:
\begin{eqnarray}
\dot{ \hat{\mathcal{O}}}=\frac i\hbar [\hat{\mathcal{H}}, \hat{\mathcal{O}}]+\sum_k\gamma_k\mathcal{D}[\hat{L}_k]\hat{\mathcal{O}}.
\end{eqnarray}
From here, starting with the desired operators, a full set of coupled equations of motion can be derived through commutation relations. Operators are of the same size as the density matrix, but their expectation values are single complex numbers. Upon attempting to evaluate the latter by averaging over the $N$ equations of motion, one is immediately confronted with another severe problem: each equation will have higher-order expectation values on its right-hand side. Here, `order' refers to the number of operators in the expectation value's operator product (\textit{e.g.}, $\langle a^\dagger a\rangle$ has order 2). These higher-order expectation values need to be solved for in turn, and this trend continues indefinitely. In other words, there is an infinite hierarchy of equations for the expectation values of products of operators\cite{quantumcumulants}. An approximate solution to this hierarchy can be achieved by truncation up to a given order. This can be implemented using R.~Kubo's generalised cumulant expansion, which allows for higher-order operator products to be expanded in terms of lower-order ones\cite{quantumcumulants, kubo1962generalized}.\\
\indent The joint cumulant, a general measure of the correlation of operators, of order $n$ is given by\cite{kubo1962generalized, quantumcumulants}
\begin{eqnarray}
\langle \hat{\mathcal{O}}_1...\hat{\mathcal{O}}_n\rangle_c=\sum_{p\in P(\mathcal{I})}(|p|-1)!(-1)^{|p|-1}\prod_{B\in p}\langle\prod_{i\in B}\hat{\mathcal{O}}_i\rangle,
\end{eqnarray}
where $\mathcal{I}=\{1,...,n\}$, $P(\mathcal{I})$ is the set of all partitions of $\mathcal{I}$, $|p|$ is the length of the partition $p$, and $B$ runs over the blocks of each partition. For $n=3$ this expression evaluates to
\begin{eqnarray}
\langle \hat{\mathcal{O}}_1\hat{\mathcal{O}}_2\hat{\mathcal{O}}_3\rangle_c & =
&\langle \hat{\mathcal{O}}_1\hat{\mathcal{O}}_2\hat{\mathcal{O}}_3\rangle\nonumber\\
& &-\langle \hat{\mathcal{O}}_1\hat{\mathcal{O}}_2\rangle \langle\hat{\mathcal{O}}_3\rangle-\langle \hat{\mathcal{O}}_1\hat{\mathcal{O}}_3\rangle \langle\hat{\mathcal{O}}_2\rangle\\
& &-\langle \hat{\mathcal{O}}_1\rangle\langle\hat{\mathcal{O}}_2\hat{\mathcal{O}}_3\rangle+2\langle \hat{\mathcal{O}}_1\rangle\langle\hat{\mathcal{O}}_2\rangle \langle\hat{\mathcal{O}}_3\rangle\nonumber.
\end{eqnarray}
A non-zero joint cumulant means that the three operators are statistically correlated with one another. If we instead assume that at least two of these operators are statistically independent,  as an approximation first proposed by Kubo\cite{kubo1962generalized}, the joint cumulant vanishes. Since the expectation value of the same order as the joint cumulant only appears once on the right-hand side, the equation can be rearranged in terms of lower order expectation values as
\begin{eqnarray}
\langle \hat{\mathcal{O}}_1\hat{\mathcal{O}}_2\hat{\mathcal{O}}_3\rangle =
&\langle \hat{\mathcal{O}}_1\hat{\mathcal{O}}_2\rangle \langle\hat{\mathcal{O}}_3\rangle+\langle \hat{\mathcal{O}}_1\hat{\mathcal{O}}_3\rangle \langle\hat{\mathcal{O}}_2\rangle \nonumber\\
&+\langle \hat{\mathcal{O}}_1\rangle\langle\hat{\mathcal{O}}_2\hat{\mathcal{O}}_3\rangle-2\langle \hat{\mathcal{O}}_1\rangle\langle\hat{\mathcal{O}}_2\rangle \langle\hat{\mathcal{O}}_3\rangle.
\end{eqnarray}
Such an expansion can be applied successively, leaving no operator products higher than the desired order. This entire procedure is automated by the Julia package QuantumCumulants.jl, written by Plankensteiner et al.\cite{quantumcumulants}. \\
\indent Unfortunately, for $N=10^{17}$, the number of coupled differential equations generated will still be enormous. Another drastic approximation used in quantum-mechanical simulations is called for: the mean-field approximation (MFA). The MFA was first employed in the early 20$\mathrm{^{th}}$ century by Pierre-Ernest Weiss to explain ferromagnetism\cite{weiss1907hypothese}. In this (uncontrolled) approximation, each emitter feels the same averaged effect of all other emitters\cite{mean, spohn1980kinetic}. For the sort of maser system we are considering, under the MFA, all molecules start in the ground state, and we neglect inhomogeneities in X-Z transition frequency, $g$ and Lindblad channels. Thus, every molecule has exactly the same quantum Langevin equation (though each molecule remains distinguishable quantum mechanically by virtue of its fixed, identifiable position within its solid host lattice). In QuantumCumulants.jl, we can use this symmetry to replace sums over $N$ with a factor of $N$\cite{quantumcumulants}, drastically reducing the number of equations that need to be solved.\\
\indent However, in real experiments, individual molecules \textit{do} sit at different locations in the `a.c.' magnetic field of the microwave mode and thus \textit{do} have different values of $g$. The mode's field can be determined using an electromagnetic solver for a given geometry and composition of the microwave cavity. To soften the brutality of the MFA, we can discretise the distribution in $g$ as a histogram to construct an approximate Hamiltonian arranged as a weighted sum over the bins of this histogram, where all emitters that fall within the same bin are assigned a coupling equal to the center value of their bin's range. 

\section{\label{sec:level2}Maser Assembly and Operation}
\begin{figure}[ht!]
    \centering
    % First subfigure
    \begin{subfigure}{\linewidth}
        \centering
        \includegraphics[width=\linewidth]{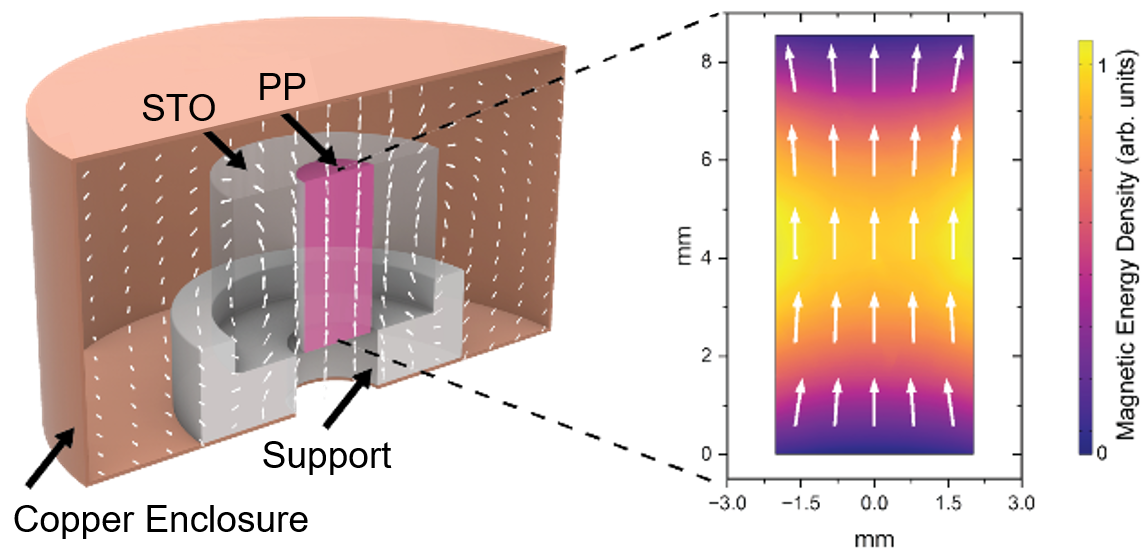}
        \caption{}
        \label{fig:schematic}
    \end{subfigure}
    % Second subfigure
    \begin{subfigure}{\linewidth}
        \centering
        \includegraphics[width=\linewidth]{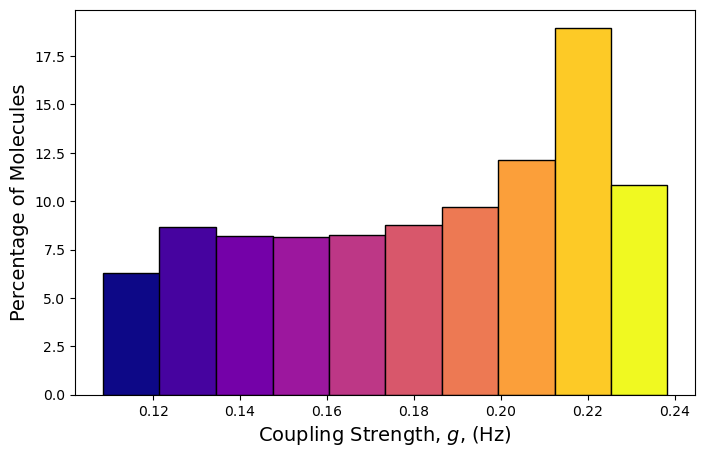}
        \caption{}
        \label{fig:distribution}
    \end{subfigure}
    \caption{(a) Rendered image of the maser cavity with magnetic vector field (white arrows) and magnetic energy density map with 
magnetic field vectors on a plane cut through the middle of 
gain medium, obtained using COMSOL's solver of Maxwell's equations. (b) A 10-bin histogram of coupling strengths obtained from calculated magnetic energy densities.}
    \label{fig:electromag}
\end{figure}

\indent The experiment simulated here is described in a previous work by Wu \textit{et al.}\cite{wu2020room}. The gain medium comprises a single crystal of \textit{p}-terphenyl, where 0.1\% of its molecules have been replaced at random by pentacene, thus forming a substitutional solid solution. As per Fig.~\ref{fig:electromag}(a), the crystal is placed in the bore of an STO dielectric ring that serves as a flux concentrator.
The ring has an inner diameter of 4.05 mm, an outer diameter of 12.00 mm, and a height of 8.60 mm. The STO ring is held 3.00 mm above a copper deck plate by a support made of cross-linked polystyrene (“Rexolite”) and placed inside a copper enclosure. The gain medium is pumped by single pulses of light from a cerium-doped yttrium-aluminium-garnet (Ce:YAG) luminescent concentrator (LC); itself pumped by a xenon flash lamp, experimentally achieving continuous masing for up to approximately 3~ms in duration\cite{wu2020room}.\\
\indent COMSOL's `RF Module' was used to obtain the magnetic energy density and field vectors for the TE$_{01\delta}$ mode in a 2.5~D, axisymmetric model of the maser's cavity.
From this, the distribution of $g$ can be found. In this simulation, the electric field of each electromagnetic mode obeys the vector Helmholtz equation: $\nabla \times \mu_r^{-1}(\nabla \times \vec{E})-k_0^2\varepsilon_r\vec{E}=0$, which comes from eliminating $\vec{H}$ in the source-free Maxwell equations in the frequency domain. A surface impedance ($\eta$) boundary condition is applied to the cavity’s internal metal walls: $\eta\hspace{0.05 cm}\hat{n}\times\vec{H}-(\hat{n}\cdot\vec{E})\hat{n}=(\hat{n}\cdot\vec{E_s})\hat{n}-\vec{E_S}$; this models energy loss without solving Maxwell's equations within the metal walls. Finally, the dielectric constant of STO was adjusted by a few percent so that the frequency of the TE$_{01\delta}$ mode exactly matched the X-Z transition,
as observed experimentally\cite{oxborrow2012room, wu2020room}.\\
\indent Fig.~\ref{fig:electromag}(a) shows the TE$_{01\delta}$ mode’s a.c.~magnetic field (white arrows), which is “funneled” through the PTP crystal, as well as the spatial variation of $g$ in the yellow-purple false-colour plot. Here $g_j$ is calculated as $g_j=\gamma\sqrt{\mu_0hf_{mode}/2V^j_{mode}}$, where $\gamma$ is the electron gyromagnetic ratio, and $V_{mode}^j$ is the effective magnetic mode volume, expressed as the ratio of the total magnetic energy stored within the cavity to the magnetic energy density at the location of the $j$th molecule\cite{breezey}:
\begin{eqnarray}
V_{mode}^j=\frac{\frac12\int \mu_0|\vec{H}(\vec{r})|^2dV}{\frac12\mu_0|\vec{H_j}(\vec{r})|^2}.
\end{eqnarray}
Fig.~\ref{fig:electromag}(b), derived from Fig.~\ref{fig:electromag}(a), shows the 10-bin histogram of $g$.

\section{Quantum Mechanics}
The Tavis-Cummings Hamiltonian, in the interaction picture, with the ensemble taken to be homogeneously resonant with the cavity mode, modified for $J$ bins of coupling, each with a central value of $g_j$ and population $N_j$, as per Fig.~2(b), is
\begin{eqnarray}
\hat{\mathcal{H}}=\sum^{J}_{j=1}\sum^{N_j}_{i=1} \hbar g_j(\hat{a}^{\dagger}\hat{\sigma}_{ji}^-+\hat{a}\hat{\sigma}_{ji}^+),
\end{eqnarray}
and the quantum Langevin equation takes the form
\begin{eqnarray}
\frac{d}{dt}\hat{\mathcal{O}}=&\frac i\hbar [\hat{\mathcal{H}}, \hat{\mathcal{O}}]+\sum_k\gamma_k\mathcal{D}[\hat{L}_k]\hat{\mathcal{O}}.
\label{eq:clustered_qLe}
\end{eqnarray}
The full set of the $k$ Lindblad channels used is listed below; they correspond, respectively, to photo-excitation and fluorescence, intersystem crossing, spin-lattice relaxation, spin dephasing, and cavity damping:
\begin{gather}
  \sum^J_{j=1}(\xi \mathcal{D}[\hat{\sigma_j}^{21}]+(\xi+k_{sp})\mathcal{D}[\hat{\sigma_j}^{12}])\label{gt:photo-excitation and fluorescence}\\
 \sum^J_{j=1}\sum_{l=3,4,5}(k_{2l}\mathcal{D}[\hat{\sigma_j}^{l2}]+k_{l1}\mathcal D[\hat{\sigma_j}^{1l}])\label{gt:intersystem crossing}\\
\sum^J_{j=1}\sum_{l,m=3,4,5;l\neq m}k_{lm}\mathcal D[\hat{\sigma_j}^{lm}]\label{gt:spin-lattice relaxation}\\
\frac12\sum^J_{j=1}\sum_{l,m=3,4,5;l\neq m}\chi_{lm}\mathcal{D}[\hat{\sigma_j}^{ll}-\hat{\sigma_j}^{mm}]\label{gt:spin dephasing}\\
\kappa((n^{th}_c+1)\mathcal{D}[\hat{a}]+n^{th}_c\mathcal{D}[\hat{a}^{\dagger}])\label{gt:photon exchange}.\end{gather}
Here, $n^{th}_c=[\exp( hf_{mode}/k_BT)-1]^{-1}$ is the average number of photons in the cavity mode due to thermal equilibrium with the environment at a given temperature; in this case, \textit{T}~=~298~K\cite{wu2020room}.
We point out that equation (\ref{eq:clustered_qLe}) and channels (\ref{gt:photo-excitation and fluorescence}) through
(\ref{gt:photon exchange}) are equivalent to Eq.~4 in Ref.~\cite{bychek2021superradiant}. Note that, in contrast to their work,
we exclude inhomogeneous broadening by setting the detuning for all molecules (and their clusters) to zero. \\
\indent We are predominantly concerned with the expectation value of the photon number operator, a second-order operator-product that is proportional to the maser's measured output power. Therefore, using QuantumCumulants.jl\cite{quantumcumulants}, we perform our cumulant expansion to second-order, finding that the time evolution of the expectation value of the photon number operator obeys
\begin{eqnarray}
\frac{d}{dt}\langle \hat{a}^\dagger \hat{a} \rangle &= \sum_{j=1}^{J} i N_j g_j ( \langle \hat{a} \, \hat{\sigma}_j^{+} \rangle - \langle \hat{a}^\dagger \, \hat{\sigma}_j^{-} \rangle ) \nonumber  
\\&-(\langle \hat{a}^\dagger \hat{a} \rangle - n^{th}_c) \kappa.
\label{eq:photon_number}
\end{eqnarray}
We thereupon use QuantumCumulants.jl\cite{quantumcumulants} to generate a complete set of equations of motion, all to second-order cumulant expansion, and solve them using the Tsitouras 5/4 Runge-Kutta method available in OrdinaryDiffEq.jl as Tsit5\cite{Tsit5, ODE}. Our initial conditions comprise setting $\langle \hat{a}^\dagger \hat{a} \rangle$ at $t = 0$ equal to a thermal number of photons and all pentacene molecules starting in their ground state (S$_0$). All spin-dynamical rates are taken from Wu~\textit{et.~al.}\cite{wu2020room}; they are reproduced here in the appendix in
Table~\ref{tab:sim_params} for completeness. 
Having solved for $\langle\hat{a}^{\dagger}\hat{a}\rangle(t)$, the maser output power is calculated as $P_{maser}=\hbar \omega_{mode}\langle\hat{a}^{\dagger}\hat{a}\rangle\kappa k/(1+k)$\cite{breezey}, where $k$ is the coupling coefficient of the cavity output, which we have set equal to 1 (corresponding to the so-called critical coupling). 

\section{Comparison to Experiment}
In our experiment, single pulses from a xenon lamp, each 3 ms in duration and with a fluence of 2.07 J/cm$^2$, are shone onto the receiving face of a Ce:YAG luminescent concentrator (LC)\cite{wu2020room}, whose area is approximately 1.7 cm$^2$. Based on measurements taken in an earlier experiment, around 17\% of each pulse's optical energy received by the LC is converted to a longer wavelength and transferred (as green-yellow light) to the PTP target. We estimate an optical pumping power, $P$, of $\approx150$ W. From this power, the rate of pump photons received by the PTP per second, $\xi$, can be calculated as
\begin{eqnarray}\xi=\frac{\lambda_p\sigma_{\lambda_p}}{hcA_p}P,\end{eqnarray}
where $\lambda_p= 592$~nm is the resonant wavelength of the singlet excitation, $\sigma_{\lambda_p}=2^{-21}$ m$^2$ is the absorption cross-section of molecular pentacene in the gain medium\cite{nelson81},
and $A_p =1.9^{-6}$ m$^2$ is the effective pump beam area in the medium.
\begin{figure}[ht!]
    \centering
    % First subfigure
    \begin{subfigure}{\linewidth}
        \centering
        \includegraphics[width=\linewidth]{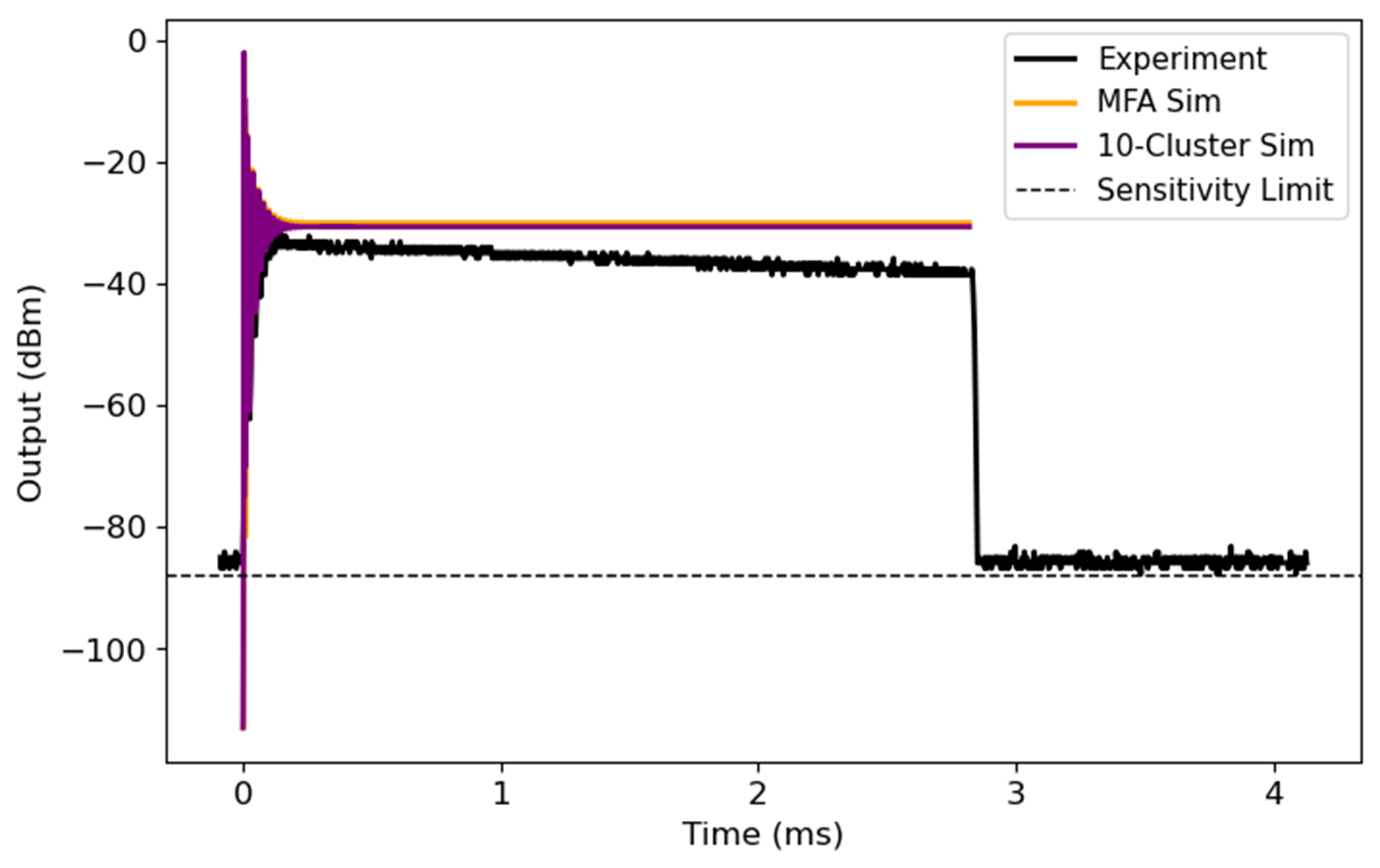}
        \caption{}
        \label{fig:schematic2}
    \end{subfigure}

    % Second subfigure
    \begin{subfigure}{\linewidth}
        \centering
        \includegraphics[width=\linewidth]{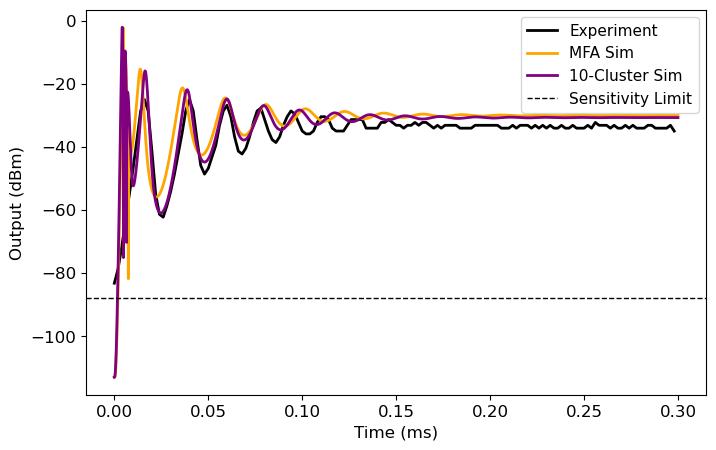}
        \caption{}
        \label{fig:distribution2}
    \end{subfigure}

    \caption{(a) Maser output power at 1.4495 GHz as a function of time for the experiment, recorded using a digital storage oscilloscope at a resolution bandwidth of 100 MHz, overlaid with the results of an MFA simulation and our 10-bin simulation (b) Output for the first 0.30 ms highlighting collective Rabi oscillations.}
    \label{fig:maseroutput}
\end{figure}
The maser’s oscillatory output at 1.4495 GHz was detected by a logarithmic detector (Analog Devices AD8318) and recorded using a digital storage oscilloscope (Tektronix TBS 1102B-EDU, 2-gigasamples/s sampling rate, 100 MHz bandwidth). \\
\indent Fig.~3 shows the fits of both an MFA and our 10-bin simulation to the experimental data. Each fit was obtained by adjusting just two unknown parameters: the number of pentacene molecules coupled to the cavity mode, $N$, and the dephasing rate,
$\chi=2/T_2^*=\chi_{lm}$ for all $l,m$.
Though there are more than $10^{17}$ pentacene molecules within the cavity's PTP crystal (as inferred from the weight of the ingredients filling the Bridgman growth vial), our best fits are obtained with
$N=2.3\times10^{15}$ used for the MFA simulation and $N=2.7\times 10^{15}$ for the 10-bin model.
We found that dephasing rates (for T$_1$'s X-Z transition) of 1.1 MHz and 0.84 MHz for the MFA and 10-bin models,
respectively, gave the best fits. These values agree well with the value of 1.1 MHz found in Ref.~\cite{wu2020room}.\\
\indent Fig.~\ref{fig:maseroutput} shows that both the MFA and 10-bin simulations can reproduce the initial collective Rabi oscillations observed before reaching steady-state. The 10-bin simulation provides a better match to both the amplitude and frequency of oscillations, with an R$^2$ value of 0.77, compared to 0.27 for the MFA model, from 0.01 to 0.20 ms.
However, both fall out of phase with the experimental data.
Once the initial oscillations have decayed, the simulated output in both our model remains fixed, whereas in the actual experiment,
the output is observed to decline slowly over time.
We speculatively attribute this decline to heating effects that our models do not account for.
We attempt to quantify the degree and consequences of optical heating in the next paragraph. \\
\indent Assuming that stimulated emission/absorption across S$_1$ and S$_0$ is weak, a pentacene molecule that is optically excited into S$_1$ has
a $p = 0.275$ probability\cite{takeda2002zero} of fluorescing straight back to the ground state S$_0$, releasing its energy optically,
and a $(1-p)$ probability of transferring by ISC into the triplet manifold, whereupon almost all of its energy is converted into heat by way of non-radiative decay\cite{clarke1976triplet}, first between T$_2$ and T$_1$ and subsequently between T$_1$ and S$_0$. 
The heat capacity of the PTP is approximately 280 JK$^{-1}$mol$^{-1}$\cite{pentaceneheat}
with the crystal containing $4.4\times10^{-4}$ mol\cite{wu2020room}.
Therefore, 3 ms of pumping with 150 W of optical power should raise the temperature of the PTP by approximately 2.6 K. 
Fig.~4 of Lang~\textit{et~al.}\cite{lang2007} indicates that the temperature sensitivity of the X-Z transition at room temperature
is approximately -0.08 MHz/K, resulting in a -0.2~MHz shift in frequency.
Over a longer timescale,
this thermal energy will also be transferred to the temperature-sensitive STO ring, the resonant frequency of which is estimated to have a temperature dependence of 2.6 MHz/K, based on the Curie-Weiss dependence of its dielectric constant\cite{STOtemp, wu2020room}. A maximum temperature increase of 0.2~K is possible for this resonator, given that the ring has a heat capacity of 100 JK$^{-1}$mol$^{-1}$ and contains $2.3\times10^{-2}$ mol\cite{STOheatcapacity, wu2020room}. This would correspond to a change in its resonant frequency of 0.3 MHz and a maximum possible frequency detuning of $\Delta f_{max}=0.3-(-0.2)=0.5$ MHz.
Given the transition's nominal inhomogeneous linewidth $= 1/(\pi T_2^*) = \chi/(2 \pi) = 0.2$~MHz or so\cite{wu2020room}, the frequency detuning due to such a temperature increase could significantly alter the effective number of pentacene molecules coupled to the maser mode, thus affecting the maser's output power. The output power will likely also be reduced by a decrease in the lifetime of the population inversion as temperature increases, as also observed in NV diamond masers\cite{zollitsch2023maser}.

\section{Maser Threshold Gedanken Simulations}
\begin{figure}[ht!]
    \centering
    \includegraphics[width=\linewidth]{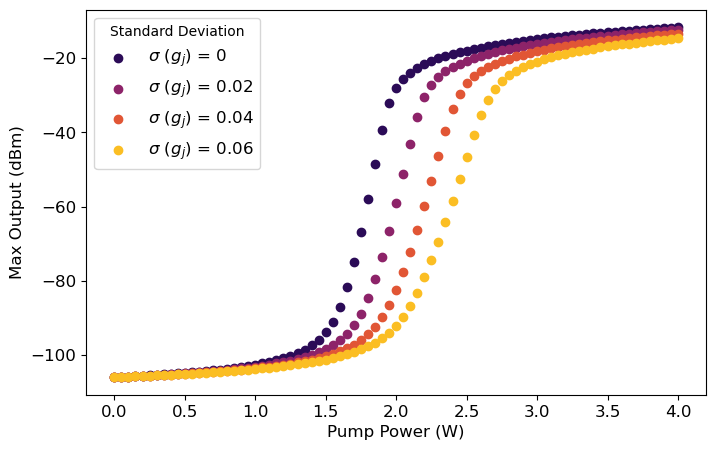}
    \caption{Maximum maser output power at 1.4495 GHz in 0.5 ms of optical pumping, against pump power, for Gaussian distributions of emitter-photon coupling strength, with $\bar g=0.18$ and a $\sigma (g_j)$ of 0, 0.02, 0.04, 0.06, respectively.}
    \label{fig:max_maser_output}
\end{figure}

In addition to fitting experimental data, we have used our beyond-MFA model to perform `Gedanken' simulations to better understand how variations in coupling strength might affect maser performance, namely the maser threshold. Here, four Gaussian distributions of varying widths (standard deviations of $\sigma(g_j)$ = 0, 0.02, 0.04, and 0.06, respectively) have been used in a 5-bin model to obtain a plot of maximum maser output (over 0.5 ms) against pumping power, $P$, as shown in Fig.~4. The mean coupling strength used was $\bar g=0.18$, taken from the distribution in Fig. 2.\\
\indent Fig.~4 shows that the pump power required to reach the maser threshold increases with $\sigma(g_j)$. Referring back to Eq.~\ref{eq:photon_number} (trivially converting from 10 to 5 bins), the time-derivative of $\langle a^\dagger a\rangle$ contains a gain and loss term. The losses, $-(\langle a^\dagger a \rangle - n^{th}_c)\ \kappa$, depend linearly on $\kappa$, a constant. The gain, $\sum_{j=1}^{J=5} i N_j g_j ( \langle a \, \sigma_j^{-} \rangle - \langle a^\dagger \, \sigma_j^{+} \rangle )$, is more complicated. The difference in the expectation values of the operator products depends, in turn, on the populations of X and Z, $\sigma_j^{XX}$ and $\sigma_j^{ZZ}$. At a given pump power $P$, each bin saturates since the population inversion is depleted by stimulated emission at the rate it is replenished by pumping, collectively giving rise to the characteristic S-shaped pumping curve observed in Fig.~4. Therefore, as $\sigma(g_j)$ increases, bins with higher $g$ saturate, limiting their contributions to the total gain, whilst bins with lower $g_j$ cannot contribute as effectively. This reduces the gain in comparison to the $\sigma(g_j)=0$ case and explains the increase in the required threshold power. Regardless of their required threshold powers, however, Fig.~4 shows that all cases tend towards the same steady-state output.

\section{Conclusions}
In summary, we highlight a more accurate quantum mechanical model of a maser. It moves beyond the typical MFA, featuring a binned distribution of couplings of emitters to the cavity mode, found through an electromagnetic field solver. Through QuantumCumulants.jl, we generated a complete set of coupled differential equations for the time evolution of the expectation values of operators. These were derived from the Quantum Langevin equation, which contained both the Tavis-Cummings Hamiltonian weighted over bins of emitters and jump operators. We solved them for an optically-pumped PTP maser using 10 bins and compared the solutions to experimental data, namely the output power over time. We had greater success in replicating quantum dynamics (collective Rabi oscillations) than the MFA, with an R$^2$ value of 0.774 compared to 0.265; however, neither model accounts for time-dependent heating and its effect on the oscillation frequency and the steady-state.\\
\indent We note here that a similarly large reduction in the effective number of molecules participating in the maser process was needed to fit the data measured in Ref.~\cite{wu2020room}. 
Though several potential culprit(s) can be speculated about [\textit{e.g.}, poorer-than-expected optical coupling due to scattering, a lower-than-calculated concentration of pentacene in the crystal due to chemical reactions and/or zone refining (here regarding pentacene as an impurity) during crystal growth, or the loss of isolated = solvated (and thus `active')  pentacene due to dimerisation/nano-aggregation], the seemingly low participation fraction remains inadequately explained. \\
\indent Having validated the model, we explore how quantum effects affect maser performance. Namely, we constructed artificial Gaussian distributions of coupling with varying widths to study the effects on maser dynamics. As the standard deviation was increased from 0, an increase in the required pumping power was observed, which is a consequence of the saturation of more strongly coupled bins. Despite this, the steady-state output power remained robust against the spreading out of the spin-photon coupling.

\hspace{1 cm}
\section{Acknowledgments}
N.R.C.~acknowledges support from the Imperial College London Department of Materials' Bursary for the Undergraduate Research Opportunities Programme.
H.W.~acknowledges support from the National Natural Science Foundation of China (Grant No. 12204040 and 12574382)
and the Beijing Institute of Technology Research Fund Program for Young Scholars (Grant No. XSQD-6120230016).
M.O.~acknowledges support from the EPSRC’s New Horizon’s Grant EP/V048430/1, “Tiger in a Cage”.

\section{Appendix}

\begin{table}[ht!]
    \centering
    \caption{Parameters used in the simulations, taken from the quasi-continuous-wave experiment for a pentacene-doped \textit{para}-terphenyl maser by Wu et al.\cite{wu2020room}.}
    \begin{tabular}{ccc}
         Rate&  Symbol& Value\\\midrule
         Cavity Mode Decay&  $\kappa$& 2.5 MHz\\
         Stimulated Emission from $S_1$&  $k_{sp}$& 42.0 MHz\\
         Intersystem Crossing&  $k_{23}$& 5.5 MHz\\
         &  $k_{24}$& 11.0 MHz\\
         &  $k_{25}$& 52.4 MHz\\
         &  $k_{31}$& 2.0 kHz\\
         &  $k_{41}$& 14.0 kHz\\
         &  $k_{51}$& 22.0 kHz\\
         Spin-Lattice Relaxation&  $k_{53}=k_{35}$& 11.0 kHz\\ 
         &  $k_{54}=k_{45}$& 4.0 kHz\\
         &  $k_{43}=k_{34}$& 28.0 kHz\\ 
    \end{tabular}
    \label{tab:sim_params}
\end{table}

The data that support the findings of this article are openly available \cite{data}. 

\bibliography{references}  

\hspace{1cm}

\end{document}